\documentclass[twoside,twocolumn,9pt]{article}
\usepackage{extsizes}
\usepackage[super,sort&compress,comma]{natbib} 
\usepackage[version=3]{mhchem}
\usepackage[left=1.5cm, right=1.5cm, top=1.785cm, bottom=2.0cm]{geometry}
\usepackage{balance}
\usepackage{mathptmx}
\usepackage{sectsty}
\usepackage{graphicx} 
\usepackage{lastpage}
\usepackage[format=plain,justification=justified,singlelinecheck=false,font={stretch=1.125,small,sf},labelfont=bf,labelsep=space]{caption}
\usepackage{float}
\usepackage{fancyhdr}
\usepackage{fnpos}
\usepackage[english]{babel}
\addto{\captionsenglish}{%
  
}
\usepackage{array}
\usepackage{droidsans}
\usepackage{charter}
\usepackage[T1]{fontenc}
\usepackage[usenames,dvipsnames]{xcolor}
\usepackage{setspace}
\usepackage[compact]{titlesec}
\usepackage{hyperref}
\usepackage{amsmath,amssymb}
\usepackage{color}
\usepackage{textcomp}
\usepackage{float}

\usepackage[utf8]{inputenc}
\usepackage[T1]{fontenc}
\usepackage{etoolbox}

\usepackage{epstopdf}

\usepackage{placeins}

\definecolor{cream}{RGB}{222,217,201}

\begin{document}

\pagestyle{fancy}
\thispagestyle{plain}
\fancypagestyle{plain}{
\renewcommand{\headrulewidth}{0pt}
}

\makeFNbottom
\makeatletter
\renewcommand\LARGE{\@setfontsize\LARGE{15pt}{17}}
\renewcommand\Large{\@setfontsize\Large{12pt}{14}}
\renewcommand\large{\@setfontsize\large{10pt}{12}}
\renewcommand\footnotesize{\@setfontsize\footnotesize{7pt}{10}}
\makeatother

\renewcommand{\thefootnote}{\fnsymbol{footnote}}
\renewcommand\footnoterule{\vspace*{1pt}%
\color{cream}\hrule width 3.5in height 0.4pt \color{black}\vspace*{5pt}} 
\setcounter{secnumdepth}{5}

\makeatletter 
\renewcommand\@biblabel[1]{#1}            
\renewcommand\@makefntext[1]%
{\noindent\makebox[0pt][r]{\@thefnmark\,}#1}
\makeatother 
\renewcommand{\figurename}{\small{Fig.}~}
\sectionfont{\sffamily\Large}
\subsectionfont{\normalsize}
\subsubsectionfont{\bf}
\setstretch{1.125} 
\setlength{\skip\footins}{0.8cm}
\setlength{\footnotesep}{0.25cm}
\setlength{\jot}{10pt}
\titlespacing*{\section}{0pt}{4pt}{4pt}
\titlespacing*{\subsection}{0pt}{15pt}{1pt}

\fancyfoot{}
\fancyfoot[LO,RE]{\vspace{-7.1pt}\includegraphics[height=9pt]{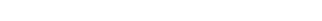}}
\fancyfoot[CO]{\vspace{-7.1pt}\hspace{13.2cm}\includegraphics{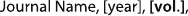}}
\fancyfoot[CE]{\vspace{-7.2pt}\hspace{-14.2cm}\includegraphics{head_foot/RF}}
\fancyfoot[RO]{\footnotesize{\sffamily{1--\pageref{LastPage} ~\textbar  \hspace{2pt}\thepage}}}
\fancyfoot[LE]{\footnotesize{\sffamily{\thepage~\textbar\hspace{3.45cm} 1--\pageref{LastPage}}}}
\fancyhead{}
\renewcommand{\headrulewidth}{0pt} 
\renewcommand{\footrulewidth}{0pt}
\setlength{\arrayrulewidth}{1pt}
\setlength{\columnsep}{6.5mm}
\setlength\bibsep{1pt}

\makeatletter 
\newlength{\figrulesep} 
\setlength{\figrulesep}{0.5\textfloatsep} 

\newcommand{\topfigrule}{\vspace*{-1pt}%
\noindent{\color{cream}\rule[-\figrulesep]{\columnwidth}{1.5pt}} }

\newcommand{\botfigrule}{\vspace*{-2pt}%
\noindent{\color{cream}\rule[\figrulesep]{\columnwidth}{1.5pt}} }

\newcommand{\dblfigrule}{\vspace*{-1pt}%
\noindent{\color{cream}\rule[-\figrulesep]{\textwidth}{1.5pt}} }

\makeatother

\twocolumn[
  \begin{@twocolumnfalse}
{\includegraphics[height=30pt]{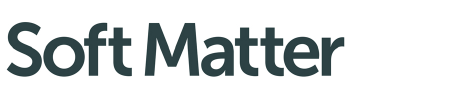}\hfill\raisebox{0pt}[0pt][0pt]{\includegraphics[height=55pt]{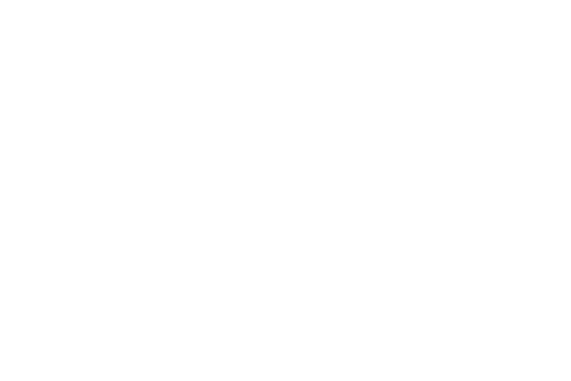}}\\[1ex]
\includegraphics[width=18.5cm]{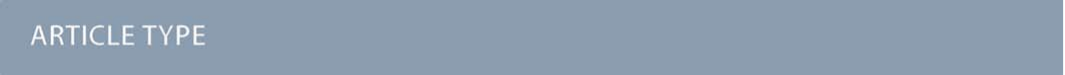}}\par
\vspace{1em}
\sffamily
\begin{tabular}{m{4.5cm} p{13.5cm} }

\includegraphics{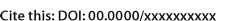} & \noindent\LARGE{\textbf{Pinch-point Singularities in Stress-Stress Correlations Reveal Rigidity in Colloidal Gels}} \\
\vspace{0.3cm} & \vspace{0.3cm} \\

 & \noindent\large{Albert Countryman,$^{\ast}$\textit{$^{a}$} H. A. Vinutha,\textit{$^{b\ddag}$},Fabiola Diaz Ruiz, \textit{$^{b\ddag}$}, Xiaoming Mao,\textit{$^{c\ddag}$} Emanuela Del Gado ,\textit{$^{b\ddag}$} and Bulbul Chakraborty \textit{$^{d}$}} \\

\includegraphics{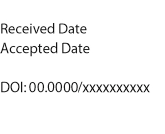} & \noindent\normalsize{We demonstrate that the spatial correlations of microscopic stresses in 2D model colloidal gels obtained in computer simulations can be quantitatively described by the predictions of a theory for emergent elasticity of pre-stressed solids (vector charge theory). By combining a rigidity analysis with the characterization provided by the stress correlations, we show that the theoretical predictions are able to distinguish rigid from floppy gels, and quantify that distinction in terms of the size of a pinch-point singularity emerging at large length scales, which, in the theory, directly derives from the constraints imposed by mechanical equilibrium on the internal forces. We also use the theoretical predictions to investigate the coupling between stress-transmission and rigidity, and we explore the possibility of a Debye-like screening mechanism that would modify the theory predictions below a characteristic length scale.}

\end{tabular}

 \end{@twocolumnfalse} \vspace{0.6cm}

  ]

\renewcommand*\rmdefault{bch}\normalfont\upshape
\rmfamily
\section*{}
\vspace{-1cm}


\footnotetext{\textit{$^{a}$~Department of Physics and Astronomy, Univerity of California Los Angeles, Los Angeles CA,
USA; E-mail:albertcoun@ucla.edu}}
\footnotetext{\textit{$^{b}$~ Department of Physics, Institute for Soft Matter Synthesis and Metrology, Georgetown University, Washington, DC, USA}}
\footnotetext{\textit{$^{c}$~ Department of Physics, University of Michigan, Ann Arbor MI, USA}}
\footnotetext{\textit{$^{d}$~ Martin Fisher School of Physics, Brandeis University, Waltham MA, USA }}





\section{Introduction}

Soft particulate gels are found in a wide range of materials from consumer products to bio-manufacturing. They are made of particulate matter (polymers, colloids, proteins, etc.) aggregated in a solid matrix, which is embedded in a fluid and typically both sparse and porous \cite{DelGadoBookChapter2016}. The soft and adaptable nature of these gel structures make them materials that can be easily manipulated in multiple ways from squeezing, stretching, spreading on surfaces and 3D printing \cite{tadros2015interfacial,petekidis_wagner_2021,cao2020design}. However, process-control is very challenging because of our lack of understanding of the mechanistic link between structure and function of soft gels. A complex interplay between particle cohesion and external driving determine their stress response \cite{aime2018microscopic,filiberti2019multiscale,colombo2014stress,koumakis2015tuning,gibaud2022nonlinear}. Additionally, the solidification process, which occurs in non-equilibrium conditions, can lead to frozen-in stresses and history dependent stress response \cite{alexander1998amorphous,anderson2002insights,royall2021real,mao2009soft,divoux2013rheological,bouzid2017elastically}. All these factors make it particularly challenging to develop theoretical frameworks that can capture the collective behavior of the constituent particles giving rise to its macroscopic emergent behavior.

Within the efforts to elucidate the rigidity of a broad range of amorphous solids, including soft particulate gels, rigidity percolation has emerged in recent years as the main theoretical framework 
\cite{behringer2018physics,henkes2022rigidity,vinutha2019force,zhang2019correlated,tong2020emergent,dashti2023emergence}, based on the idea that locally rigid structures (due to mechanical constraints such as contacts, chemical
bonds or steric repulsion) percolate through the material. This  concept has been pivotal in clarifying the role played by microscopic properties in giving rise to macroscopic mechanical response, that is, which parts of the microstructure, self-assembled during the solidification process,  may satisfy, and how, the conditions corresponding to mechanical equilibrium and Maxwell's criterion for rigidity. 

In the context of granular solids, it has become increasingly evident that stress 
correlations emerging from the constraints of mechanical equilibrium are anisotropic, sub-dimensional and long-ranged \cite{behringer2018physics,nampoothiri2020emergent,vinutha2019force}. The theoretical framework developed by Nampoothiri {\it et al.} \cite{nampoothiri2020emergent,nampoothiri2022tensor} has demonstrated a rigorous mapping of a tensor gauge theory -- a Vector Charge Theory (VCT) similar to field theories for quantum spin liquids \cite{pretko2017generalized} -- to mechanics of amorphous solids, formulated in terms of mechanical constraints. 
The mapping leads to an ``emergent elasticity" and an accurate description of stress localization in granular solids, predicting quantitatively the outcomes of experiments with photoelastic grains and computer simulations of model granular solids. The VCT mapping proves that those specific features of the stress correlations, i.e. being anisotropic, sub-dimensional, and long-ranged,  emerge purely from the constraints, and dictate mechanical response and elasticity.  This aspect could not be captured by  the rigidity percolation framework and has relevance well-beyond granular solids, since it applies broadly to athermal amorphous solids. Soft particulate gels are a class of materials where thermal fluctuations are usually important drivers of self-assembly and solidification, however high interaction strengths are required to form rigid, stress-bearing structures that control both dynamics and elasticity \cite{whitaker2019colloidal,keshavarz2021time,zhang2019correlated,hsiao2012role,bantawa2023hidden,bouzid2017elastically,fenton2023minimal}. Interestingly, it has been shown that spatial correlations of stress fluctuations measured in model particulate gels through computer simulations have the same characteristics predicted by the VCT mapping \cite{vinutha2023stress}, indicating that also in these very soft materials the constraints imposed by mechanical equilibrium manifest themselves in the same way. These findings suggest that there may be a connection between the rigidity percolation framework and the VCT approach. Establishing such  a connection and clarifying which aspects of soft-particulate-gels rigidity and mechanics can be quantified by the VCT framework could allow for  a deeper understanding of the connection between these two different theoretical frameworks.  Further, it can clarify  similarities or differences across complex amorphous materials.     

Here, we wish to contribute to some of these questions and explore the implications of the VCT paradigm for soft particulate gels. We use the predictive framework of VCT to interpret the stress-stress correlations measured in model colloidal gels through 2D computer simulations and show consistency between the rigidity percolation framework and VCT framework in inferring the rigidity onset. From the stress correlations, we also estimate the emergent elastic modulus of gels as predicted by VCT, which reveals correlations between normal stresses and mechanical strength in this class of materials. Finally, the comparison between the VCT predictions and the simulations motivates us to extend the VCT framework by incorporating a length scale, associated with stress fluctuations, which is absent in the case of granular solids. 

The paper is organized as follows: Section 2 briefly reviews the basic ingredients of the VCT framework and its predictions for the stress–stress correlations in rigid gels. Section 3 contains information on the numerical model and simulations used. We then discuss the results in Sec. 4 and provide a summary and outlook in Sec. 5.

\section{Theoretical Framework} 
\label{theory-simulations}
\subsection{Tensor gauge theories} 

Theories of emergent elasticity of pre-stressed solids have been recently mapped on to $U(1)$ Tensor Gauge Theories~\cite{pretko2017generalized}. These continuum theories include residual or frozen stresses (pre-stress) that appear as  a consequence of either (i) presence of external body forces such as gravity, hydrodynamic drag or active forcing~\cite{nampoothiri2022tensor,nampoothiri2020emergent}, or (ii) as a result of relieving geometric frustration~\cite{Livne:2023aa,lemaitre2021anomalous,Kumar:2022aa,fu2024long}. The prestress appears as the analogs of the familiar polarization field  present in a dielectric in standard electrostatics. Dielectric screening screens out unbound charges. In contrast to standard electrostatics, however, this polarization field is a symmetric second rank tensor.  These theories also accommodate a Debye type screening that occurs, as in electrostatics, via free charges and where a screening length emerges~\cite{Livne:2023aa,lemaitre2021anomalous,Kumar:2022aa,fu2024long,Surajit_Kabir}.  

  
In tensor gauge theories, where the observable fields are {\it symmetric $2$nd rank tensors, two distinct  Gauss's laws are allowed, one with vector charges (VCT) and the other with scalar charges (SCT)~\cite{pretko2017generalized}. In mapping to elasticity theories, the Gauss's law applies to the stress tensor.}  

The Gauss's law in a VCT based theory is $\partial_i \sigma_{ij} = f_{j}^{ext}$, reflecting momentum conservation. The vector ``charges'' in the theory are the externally exerted forces, either boundary or body forces. This is the basis of the theory that maps VCT to  mechanics (VCTG) discussed further in the next section. 
SCT gauge theories emerge via a {\it duality} mapping of defects in solids to scalar charges in the Gauss's law: $\partial_i \partial_j \tilde{\sigma}_{ij} = \rho$. These are theories that live in two-dimensions, and $\tilde{\sigma}_{ij} = \epsilon_{ia} \epsilon_{jb} \sigma_{ab}$. This mapping has been extensively applied to defects in crystalline solids~\cite{Pretko_Radzihovsky}, where the scalar charges represent disclinations, and bound dipoles represent dislocations. The adaptation of this type of theory to amorphous solids has been based on the crucial finding that the lowest-order ``defects'' in such solids are bound quadrupoles of the scalar charges of SCT~\cite{Livne:2023aa}. A non-zero value of $\rho$ indicates an incompatible stress~\cite{Livne:2023aa}.

In this paper, we use the VCTG framework to analyze stress-stress correlations in gels undergoing a floppy to rigid transition through numerical simulations of dynamical arrest, as deduced from a rigidity percolation analysis \cite{zhang2019correlated}.







\subsection{VCT-based Elasticity Theory}

In VCTG 
the pre-stresses existing in a granular solid or other amorphous solids can be described as the polarization field in a dielectric, in the electrostatic limit of VCT~\cite{nampoothiri2020emergent,nampoothiri2022tensor,Surajit_Kabir}. 
 Crucially, VCTG emerges upon ensemble averaging over all states that are subjected to the same set of external forces, $\bf{f}^{ext}$. In jammed granular solids, for example, this translates to the externally imposed shear stress and/or pressure. The reasoning, which is familiar in the context of many frustrated systems~\cite{spin_liquids_review}, is that there is no unique reference state and states that obey the constraint of force balance (Gauss's law) are equiprobable.   

VCTG is a stress-only formulation of elasticity theory: there is no unique reference state or metric, and therefore no concept of strain. In contrast, as in an electrostatic dielectric, there is an unscreened stress field, $\hat{E}$, and the screened stress field is $\sigma_{ij} = E_{ij} + P_{ij}$, where $\hat{P}$ is the induced polarization field. In the linear dielectric version of VCTG,  $\hat{P} = \mathcal{\chi} \hat E$, or equivalently, $\hat{\sigma}=\Lambda^{-1} \hat{E}$, where $\Lambda^{-1}$ is a $4$th rank tensor that appears as a coupling constant in the field theory and gives rise to the emergent elastic moduli~\cite{nampoothiri2020emergent,nampoothiri2022tensor}. The picture is that a stress-bearing structure emerges as a consequence of satisfying Gauss's law with given external forces. The {\it compatibility} relation in VCTG elasticity appears as the static limit of the analog of Faraday's law: $\nabla \times \nabla \times E =0$. There is no such condition imposed by VCTG on the induced stress, $\hat{P}$. In a linear dielectric, $\hat{P}$ is slaved to $\hat{E}$, however VCTG does allow for incompatibility in $\hat{P}$ within a nonlinear dielectric theory.

VCTG is a field theory that predicts response and correlation functions, and it provides a rigorous basis for the emergence of stress heterogeneities such as ``force-chains''~\cite{nampoothiri2020emergent,nampoothiri2022tensor,Surajit_Kabir,vinutha2023stress}. In this paper, we test the applicability of the {\it linear dielectric} theory across the rigid-floppy transitions. We also look for the possible emergence of a Debye-type screening length in the stress-stress correlations.  

\subsection{Pinch-point Singularities}

\section{Numerical Simulations}
Gel configurations in two dimensions are obtained using a 2D model of $N(=10000)$ interacting monodisperse colloidal disks that undergo gelation as described in Refs.\cite{zhang2019correlated}. The particles interact via a short range attractive potential 
$
U(r) = A \epsilon \left(a\left( d/r \right)^{18} - \left(d/r\right)^{16}\right)$   
where $r$ is the interparticle distance, $d$ is chosen as the particle diameter, $U_{0}$ and $a$ are dimensionless constants. The values of $U_{0}=6.27$ and $a=0.85$ are chosen to obtain a short-ranged attractive well of depth $\epsilon$ and range $r_c \approx 0.3 d$ (the potential is cut and shifted to $0$ at large distances). 
We perform Molecular Dynamics (MD) simulations in a square box of size $L$ with periodic boundary conditions. To prepare the gels, we first reach thermal equilibrium at a high temperature and then slowly quench the particle configurations to a target temperatures $T$ using a Nos\'{e}--Hoover thermostat, allowing the particles to aggregate and form gel networks \cite{noro2000extended,colombo2014self}. The mechanical equilibrium in the final gel state is obtained by slowly withdrawing all kinetic energies with overdamped dynamics, following the protocol described in Refs. \cite{zhang2019correlated}, which mimics the slow aging typical of these systems \cite{colombo2014stress,bantawa2021microscopic}.  

Once all kinetic energy has been withdrawn, the stress tensor $\boldsymbol{\hat{\sigma}}$ is computed for each configuration using the following equation:
\begin{equation}\label{FMpar}
\boldsymbol{\hat{\sigma}} = \frac{1}{A}\sum_{\rm~p}\sum_{\rm~s>p} \boldsymbol{r}_{\rm~p,s} \otimes  \boldsymbol{f}_{\rm~p,s}
\end{equation} 
Where $A$ is the simulation box area, $\boldsymbol{r}_{\rm~p,s}$ is the vector connecting the center of particles \textrm{p} and \textrm{s}, and $\boldsymbol{f}_{\rm~p,s}$ is the force between them. The index \textrm{s} iterates over the neighbors of particle \textrm{p} within the interaction range  $r_{\rm~p,s} \le r_c$. We obtain the pressure $P$ from the trace of the stress tensor. 
Following Ref. \cite{vinutha2023stress}, we compute the stress correlations in Fourier space using,
\begin{eqnarray}\label{formulacorel}
C_{ijkl} (\boldsymbol{q}) = \langle \Delta \sigma_{ij}(\boldsymbol{q}) \Delta \sigma_{kl}(-\boldsymbol{q})\rangle \\
\Delta \sigma_{ij}(\boldsymbol{q}) = \sum_{\rm~p=1}^{N} (\sigma_{ij \rm~p} - \langle \sigma_{ij} \rangle) \exp(i\boldsymbol{q \cdot r}_{\rm~p})
\end{eqnarray}
 where $\boldsymbol{q}$ denotes the wave vector, $\boldsymbol{r}_{\rm~p}$ is the \textrm{p}$^{th}$ particle position vector, $\sigma_{ij \rm~p}$ denotes the $ij$-component of the stress tensor of particle \textrm{p}, and $\Delta \sigma_{ij}$ represents the fluctuation of the stress tensor with respect to the average value $\langle \sigma_{ij} \rangle$ within the corresponding individual configuration. The angular brackets $\langle ...\rangle$ denote an ensemble average over several configurations corresponding to the same target temperature $T$ used in the gel preparation and the same particle volume (surface) fraction. For each of these samples we perform an average over all the particles before taking the ensemble averages. The stress fluctuations in Fourier space, obtained through these calculations, are then coarse-grained by imposing a cutoff at large $q$, corresponding to $q_{max} = 2\pi/d$, i.e. we do not consider any stress fluctuations occurring at length scales shorter than $d$. The lower $q$ cutoff in Fourier space is set by the simulation box size $q_{min} = 2\pi/L$.
 
All the simulation quantities described in the following are expressed in reduced units; $d$ is used as the unit of length, $\epsilon$ is used as the unit of energy, and $m$ (the particle mass) is used as the unit of mass. The reduced temperature is expressed in units $\epsilon/k_B$, where $k_B$ is the Boltzmann's constant. The unit of stress is therefore $\epsilon/d^2$ in 2D. We define an approximate volume (surface) fraction $\phi = \frac{\pi d^2 N}{4L^2}$. All the simulations are performed using LAMMPS \cite{lammps}. All the data for the stress correlations have been averaged over more than $100$ independently generated samples. 

Following Ref. \cite{zhang2019correlated,vinutha2023stress}, for the discussion here we have selected data for the same $T=0.18 \epsilon/k_B$ and particle surface fraction $\phi=0.39,0.5$, which were located respectively in the floppy and rigid region of the rigidity phase diagram obtained for this model. A specific gel configuration is considered to be rigid if the pebble game algorithm \cite{jacobs1997algorithm} identifies a spanning rigid cluster, i.e., a network of particles where all the degrees of freedom are blocked by contacts or constraints, allowing only for overall translations or rotations of the entire network. 
A gel configuration is instead considered floppy if the corresponding spanning network of aggregated particles is not rigid according to the pebble game algorithm \cite{jacobs1997algorithm,zhang2019correlated,vinutha2019force}. The phase boundary reported by Zhang {\it et al} \cite{zhang2019correlated} for the rigidity phase diagram is based on the percolation probability measured over many independently generated samples at various $T$ and $\phi$. 

\begin{figure*}[ht!]
\centering
\includegraphics[width=\textwidth]{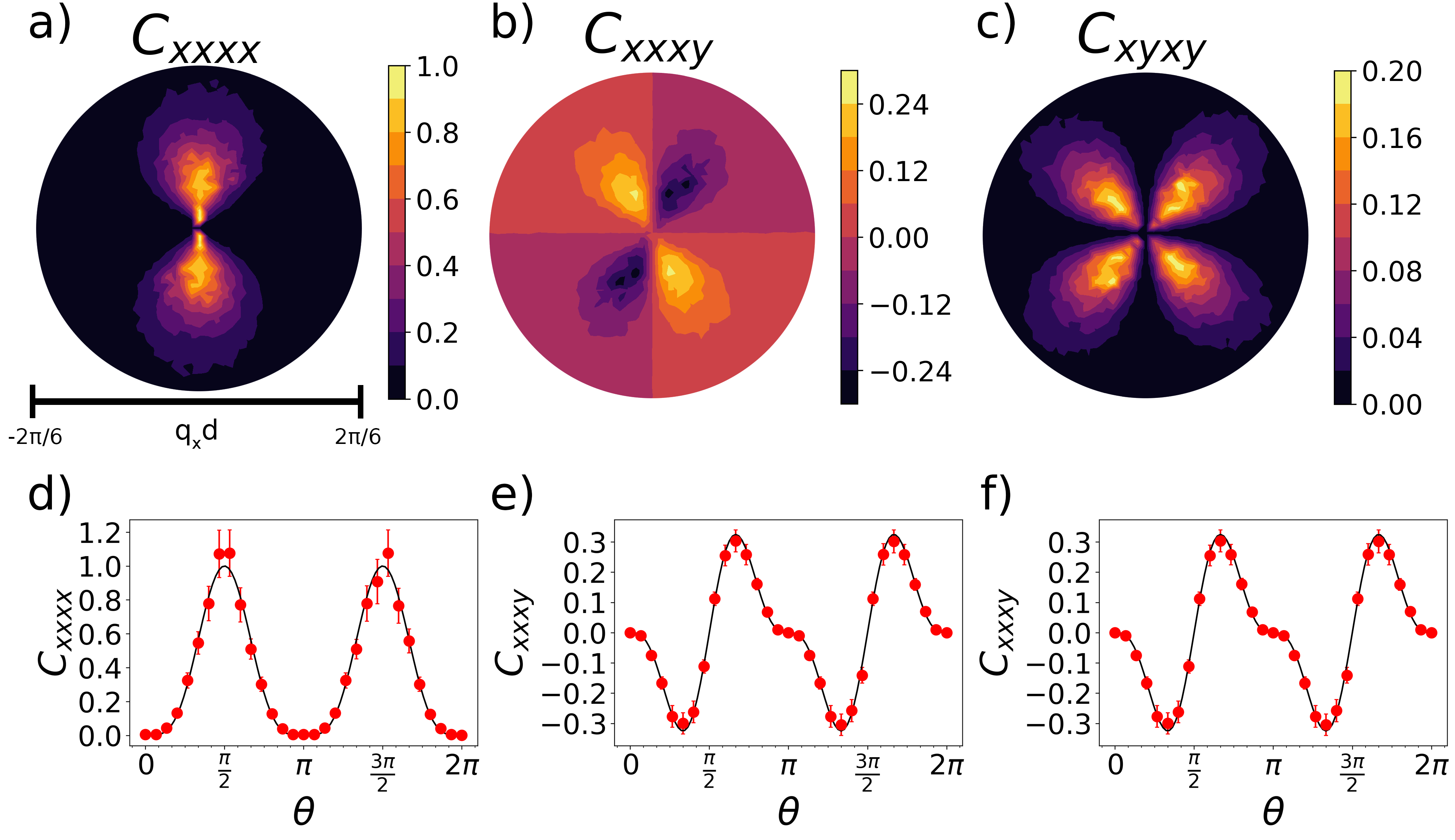}
\caption{\label{stress2D} {\bf{Components of ensemble-averaged stress-stress correlations for $\phi = 0.5$ gels.}}  Correlation intensity of the function $C_{ijkl}(q_x,q_y)$ are shown in the top row for $ijkl=xxxx$ (a), $xxxy$ (b), and $xyxy$ (c), normalized by the maximum value of $C_{xxxx}$ . The red series in the (d), (e), (f) shows the same correlation functions averaged into angular bins in Fourier space (error bars shown correspond to standard error from averaging into angular bins), with fit lines to isotropic VCTG theoretical forms shown in black. Within the set of configurations generated with these parameters, the configurations averaged here were those that exhibited rigidity percolation in two directions.}
\end{figure*}

\begin{figure*}[ht!]
\centering
\includegraphics[width=\textwidth]{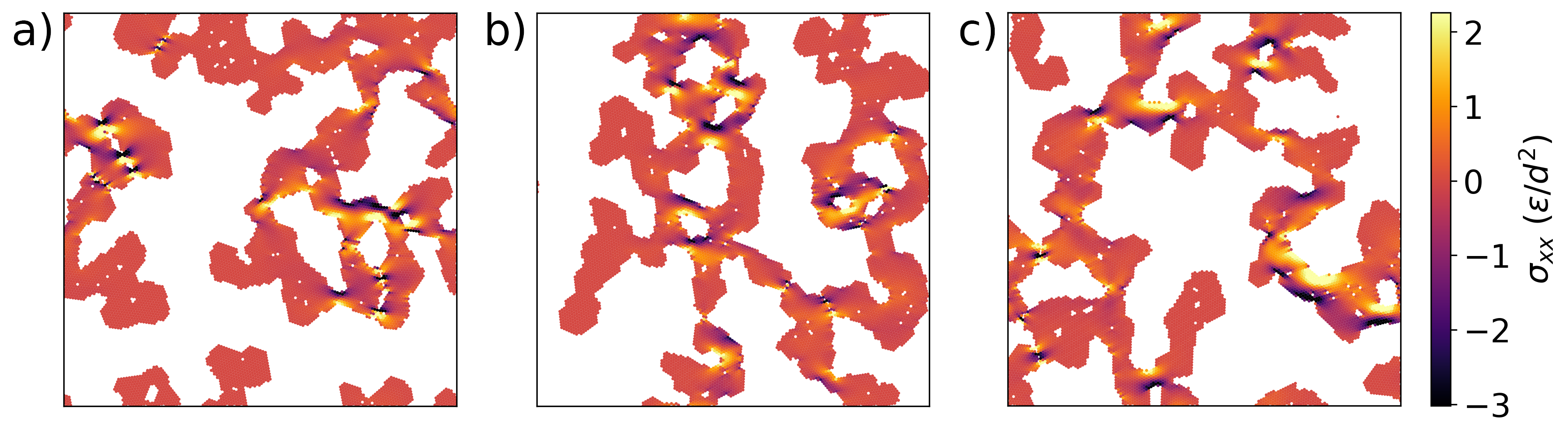}
\caption{\label{snapshots} {\bf{Snapshots of gel configurations}} which exhibit rigidity percolation in 0 directions (a), 1 direction (b), and 2 directions (c). The above gel configurations were generated with a packing fraction of $\phi=0.39$, which lies on the floppy side of the phase boundary reported by Zhang {\it et al} \cite{zhang2019correlated}. }
\end{figure*}

\section{Results}
We begin by verifying the predictions of VCTG theory for gels at $\phi=0.5$, which corresponds to the rigid region of the rigidty phase diagram \cite{zhang2019correlated}. At this high surface fraction, the gel configurations are most likely to be rigid, and hence their stress-stress correlations in Fourier space should exhibit a pinch-point structure: we should observe long-range correlations with correlation intensity that has only angular dependence. In Fig. \ref{stress2D}(a)-(c), we show selected stress-stress correlations maps of gel configurations at $\phi=0.5$. In order to produce these plots, the data are re-interpolated onto a polar grid in Fourier space with an interpolation distance of $q=1/d$. 
The maps display precisely the two- and four-fold patterns predicted by VCTG\cite{nampoothiri2020emergent} for $C_{xxxx}$, $C_{xxxy}$, and $C_{xyxy}$, respectively, as well as  the pinch-point profile at $q\approx0$. Note that in 2D we have in total six stress-stress correlations and, while we show for brevity only three of them, all of them satisfy the theoretical predictions, consistent with the results in Ref.~\cite{vinutha2023stress}. 
The angular variation of the stress correlation intensities are shown in Fig. \ref{stress2D}(d)-(f). The polar grid used in the maps allows for grid points to be allocated into angular averaging bins with even populations. The angular variation of all six components of $C_{ijkl}(\vec{q})$ measured are perfectly fit by the predictions of VCTG shown by the lines in the angular plots in Fig. \ref{stress2D}. The fit to the theory for the angular dependence of the stress correlations can be used to extract the elastic modulus of the gels, as discussed below. 

\begin{figure*}[ht!]
\centering
\includegraphics[width=\textwidth]{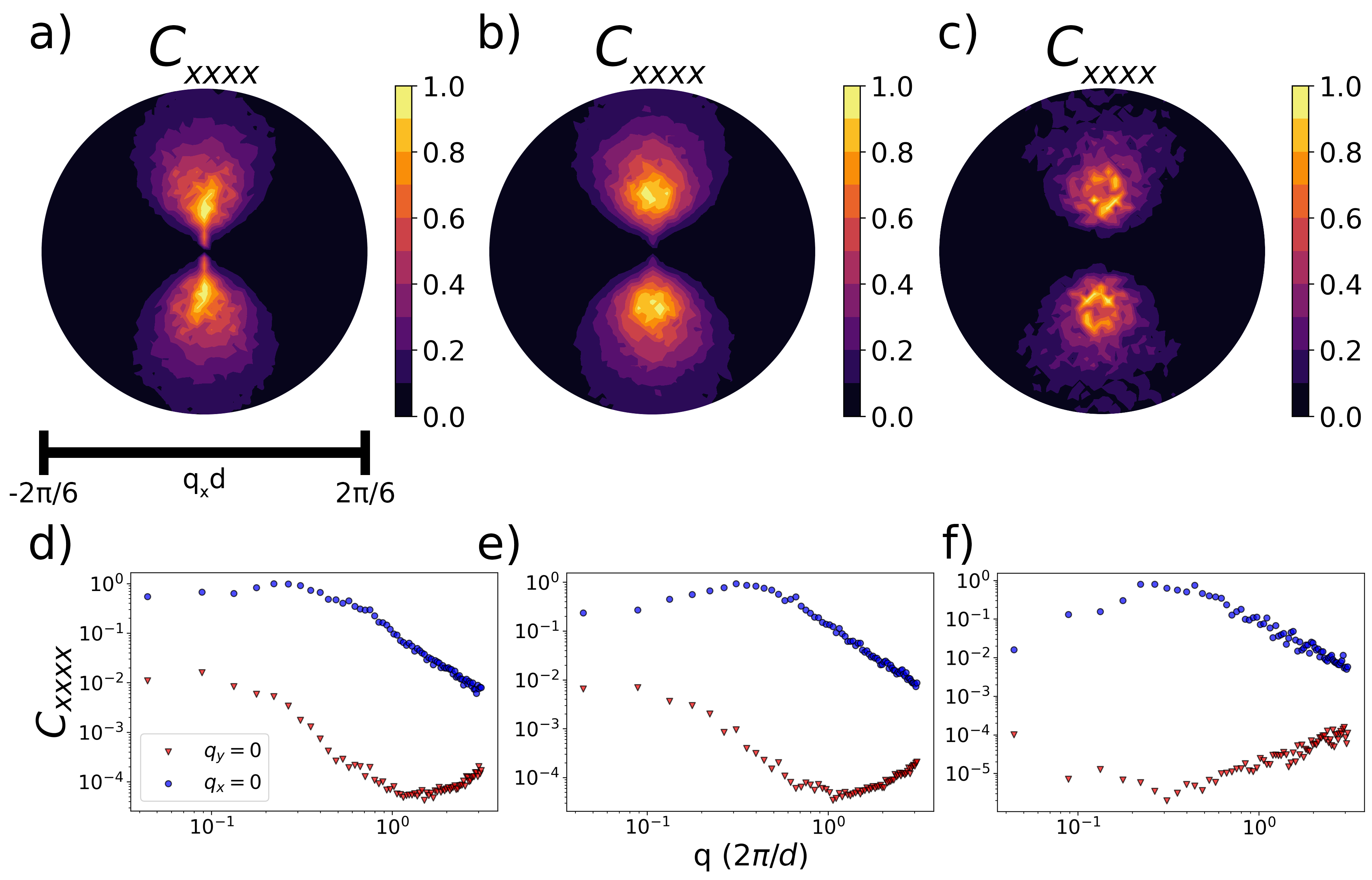}
\caption{\label{crosssection} {\bf{Analytical behavior of the pinch point singularity in $C_{ijkl}(\vec{q})$ correlates with degree of rigidity percolation.}} Spectra of $C_{xxxx}$ (normalized by the maximum value of $C_{xxxx}$) are ensemble averaged over configurations which exhibit rigidity percolation in 2 directions (a), 1 direction (b), and 0 directions (c). Cross sections of these spectra along $q_y=0$ and $q_x=0$ are shown in (d), (e), (f) in red and blue, respectively. The population of gels that did not percolate are shown to lack flat $|q|$-dependence in $C_{ijkl}(\vec{q})$, indicating that these configurations are not rigid by the reckoning of VCTG. }


\end{figure*}
 
\subsection{Rigidity Percolation and VCTG theory}
According to the rigidity percolation framework \cite{zhang2019correlated,vinutha2019force,nampoothiri2022tensor}, gel structures require a mechanically stable spanning cluster able to transmit stresses (in addition to standard connectivity percolation). The rigidity percolation analysis explicitly takes into account the gel network architecture in which mechanical constraints (local and global) operate to obtain the macroscopic emergent rigidity. Zhang {\it et al.,}\cite{zhang2019correlated} had studied the nature of the rigidity transition in our gel model and obtained the rigidity phase diagram in temperature-density plane. For the preparation temperature $T=0.18 \epsilon/k_{B}$, gels formed at both $\phi=0.39$ and $\phi=0.5$ had configurations that may have percolating rigid clusters. This fraction was significantly higher in the latter case, placing these two particle surface fractions on different sides of the rigidity boundary. A particular location on the rigidity phase diagram, in fact, is merely defined by whether it is more likely to produce a configuration that is rigid or floppy: both rigid and floppy configurations can be generated for all choices of $(\phi,T)$ near the phase boundary \cite{zhang2019correlated}.

In order to understand the role of the nature of rigid clusters on the gel elasticity and connect rigidity percolation theory and VCTG theory, we further split the gel configurations based on the degree of percolation of the largest rigid cluster, which can be 0, 1 or 2, if the largest rigid cluster is not percolating (0), percolating in only one direction (1), or in both directions (2). The latter case corresponds to a rigid network, and the former two are considered floppy. The fact that the two datasets chosen for this analysis are on opposite sides of the rigidity phase boundary \cite{zhang2019correlated} implies that the particle fraction of $\phi=0.39$ produces predominantly floppy configurations, whereas $\phi=0.5$ leads to predominantly rigid configurations. In Fig. \ref{snapshots}, we show snapshots of gels configurations at $\phi=0.39$ corresponding to no percolation (a), 1-direction percolation (b) and 2-direction percolation (c). More specifically, the low-density dataset on the floppy side of the phase diagram consists of $12$ configurations that percolate in 0 directions, $102$ configurations that percolate in 1 direction, and $65$ configurations that percolate in 2 directions, resulting in a rigidity percolation probability of $0.36$. The higher-density dataset on the rigid side of the phase diagram ($\phi=0.5$) consists of $1$ configuration that percolates in 0 directions, $47$ configurations that percolate in 1 direction, and $76$ configurations that percolation 2 directions, resulting in a rigidity percolation probability of $0.61$. 

Within VCTG, pinch-points are a qualitative indicator of rigidity: if a pinch-point in $C_{ijkl}(\vec{q})$ is present, the system should exhibit rigidity at long length scales. We are interested in understanding how this property correlates with the nature of the rigid cluster. 
Fig. \ref{crosssection} shows results for the correlation function $C_{xxxx}({\bf{q}}) \equiv \langle\sigma_{xx}({\bf{q}})\sigma_{xx}({\bf{-q}})\rangle $ for  gel configurations at $\phi = 0.39$ with the largest rigid cluster percolating along $0,1,2$ directions, respectively. For each case, we average over all available configurations with the same degree of rigidity percolation. The maps (Fig. \ref{crosssection}(a)-(c)) show the existence of the pinch-point singularity only in the case where the rigid cluster percolates in both directions. The size of the pinch-point is established by the asymptotic values of $C_{ijkl}$ as $|q| \rightarrow 0$ measured along the $q_y=0$ direction. The presence/absence of a pinch point can be therefore quantified from the values of $C_{xxxx}$ at the smallest $q$ (Fig. \ref{crosssection}(d)): a strong difference in relative value measured along the $q_x=0$ and $q_y=0$ cross-sections indicates that a pinch-point singularity is present.  
The pinch-point becomes less obvious for the gels with rigid clusters percolating along one direction, see Fig. \ref{crosssection}(b) and (e). For the gel systems where there is no percolating rigid cluster, Fig. \ref{crosssection}(c), we observe no signal at $q \approx 0$, see Fig. \ref{crosssection}(f). In other words, the pinch-point is clearly absent. The plots Fig. \ref{crosssection}(d)-(f) also show that for only fully rigid gels feature, at low $q$, the flat $q$-profile of the stress correlations predicted by VCTG. This analysis demonstrates that the pebble game evaluation of the rigidity is fully consistent with the behavior of the stress-stress correlations and of the pinch-point singularity predicted by VCTG.  

We also note that the correlations profiles in Fig. \ref{crosssection}(d)-(f) indicate the presence of a length scale beyond which the rigid or floppy behavior of the gels is clearly distinguishable, and beyond which the VCTG predictions are satisfied for the fully rigid gels. 
Stress fluctuations at smaller lenghtscales (larger $q$ values) seem to be similar across floppy and rigid gels.

Such a length scale was not detected by the pebble game analysis and may be a distinctive feature of gel structures. In granular solids, for example, VCTG applies all the way up to the grain scale~\cite{nampoothiri2020emergent,nampoothiri2022tensor,Surajit_Kabir}.

\begin{figure*}[h!]
\centering
\includegraphics[width=\textwidth]{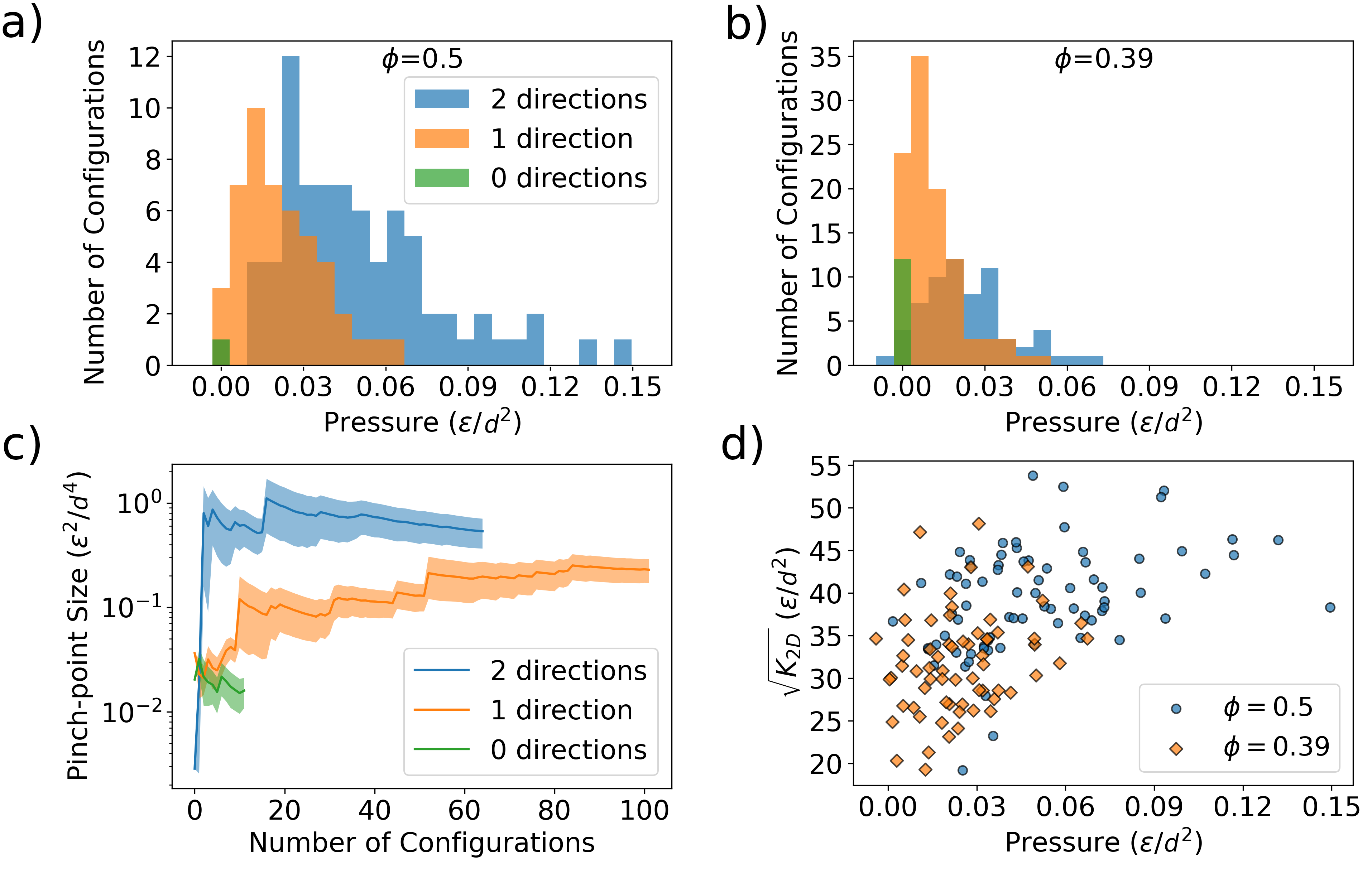}
\caption{\label{pressurestory}{\bf{Rigidity of gel configurations at particular degree of rigidity percolation correlates with configuration pressure.}} Subfigures (a) and (b) depict the pressure distributions of the $\phi=0.5$ rigid gel dataset and the $\phi=0.39$ floppy dataset, with each distribution separated into groups of equivalent rigidity percolation. The distribution associated with 2-directional percolation consists of mostly higher-pressure configurations. Configurations which do not percolate display lower total pressure, and vice versa. This correlation persists across different locations of the phase diagram, even as the underlying probability distribution affecting the probability of generating a configuration which percolates fully is changing. Subfigure (c) depicts the size of the pinch-point singularity measured in the $\phi=0.39$ dataset as additional statistics are taken into account, quantifying the disappearance of the pinch point seen in non-rigid configurations. In subfigure (d), Gel stiffness $(K_{2D})$ measured from fit is plotted against total configuration pressure. The square root of $K_{2D}$ is measured in order to have the same units as Tr$(\sigma)$. Circles are configurations from the $\phi=0.5$ dataset and diamonds are from the $\phi = 0.39$ dataset.}
\end{figure*}

\subsection{Pressure and Elastic modulus}
The natural ensemble for VCTG is a stress ensemble (which means a pressure ensemble in the case of isotropic compression). In order to better connect the predictions of the theory with properties of rigid clusters obtained instead at constant particle surface fraction, we have computed the distribution of pressure across the gel configurations produced in the simulations and investigated if and how the degree of percolation of rigid clusters can be correlated with the pressure values. Fig.\ref{pressurestory}(a) and (b) show that, regardless of the particle surface fraction, fully rigid gels whose rigid clusters percolate in both directions tend to have higher pressures. Partially floppy gels, for which a rigid cluster span the system only in one direction, tend to have lower pressures, and completely floppy gels correspond to zero pressure states. While one would expect that simply increasing the particle surface fraction may be responsible for the overall increasing pressure of rigid gels, the distinct correlations with the degree of percolation of the rigid cluster point instead to an interesting possible connection between pressure heterogeneity and rigidity percolation. The fact that these correlations may be specifically due to the rigidity (or the absence of) is confirmed in Fig. \ref{pressurestory}c), showing the correlation between the pinch-point size and the degree of percolation of the rigid cluster for the less dense gels, connecting the pinch-point size to the distribution of pressures in the same sample at the same particle surface fraction. 


\begin{figure*}[ht!]
\centering
\includegraphics[width=\textwidth]{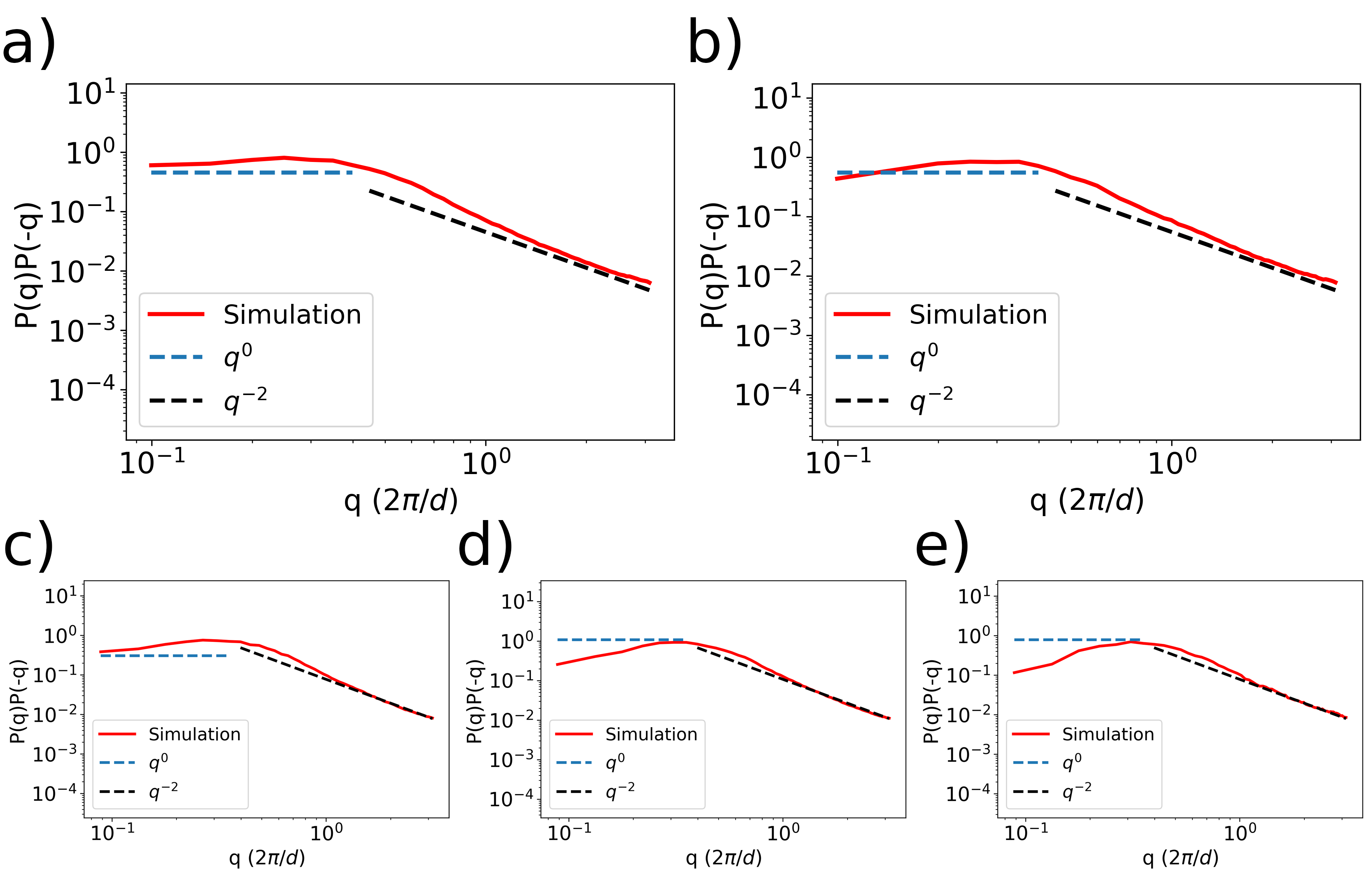}
\caption{\label{lengthscale} {\bf{Pressure-Pressure correlations reveal rigidity-independent length scale.} }Subfigures (a) and (b) depict the value of $P(q)P(-q)$ (normalized by the maximum value of $C_{xxxx}$) averaged over all angles for $\phi=0.5$, for gels with rigidity percolation in 2 directions and 1 direction, respectively. A flat power law is plotted near $q=0$ in blue, and a power law proportional to $q^{-2}$ is plotted near $qd=\pi$ in black. Subfigures (c), (d), and (e) depict the same functions for $\phi=0.39$, for gels which percolate in 2 directions, 1 direction, and 0 directions, respectively.}



\end{figure*}

\subsubsection{Elastic Moduli}
\label{results}
As discussed above, the VCTG theory provides a way to compute the gel elastic moduli from the set of stress-stress correlations, $C_{ijkl}$. As the VCTG theoretical form of $C_{ijkl}(\vec{q})$ is solely dependent upon Fourier angle, the stress correlations from each rigid gel configuration was averaged across configurations with the same particle surface fraction and with the same degree of rigidity percolation, then re-interpolated onto a polar grid in Fourier space. This allowed for $C_{ijkl}(|q|,\theta)$ to be averaged into angular bins for the purposes of fitting to the theoretical form of $C_{ijkl}(\vec{q})$ (as shown in Fig.\ref{stress2D}(d)-(f)). The nonlinear fitting package SymFit was used to simultaneously fit all 6 components of the stress-stress correlations, using an isotropic form of the VCTG elastic modulus tensor (presented here in Voigt notation), that appears as a coupling constant in the field theory:

\begin{equation*}
    \Lambda^{-1} = \begin{bmatrix}
        \lambda + 2\mu & \lambda & 0 \\
        \lambda & \lambda+2\mu & 0 \\
        0 & 0 & 2\mu
    \end{bmatrix}
\end{equation*}
\\
Although this form of elastic modulus tensor is dependent upon two Lamé coefficients $\lambda$ and $\mu$, in 2D the functions $C_{ijkl}(\vec{q})$ depend only on the stiffness parameter $K_{2D}\equiv\mu(\lambda+\mu)/(\lambda+2\mu)$~\cite{nampoothiri2020emergent}. The fitting quality ($r^2 = 0.981$) confirms the strong agreement between the angular dependence of the measured stress correlations and the one predicted by theory. 
  
Following this demonstration, $K_{2D}$ was then measured for individual rigid gel configurations at different packing fractions. By performing the fitting process on individual configurations, it is then possible to examine the makeup of the ensemble-averaged prediction of $K_{2D}$. Although this type of fitting would normally be too noisy to perform on a single configuration, the averaging into angular bins applies a degree of smoothing which enables for reasonable fits to be carried out. We note that the value of $K_{2D}$ is proportional to the magnitude of the correlation function, and hence determining a value for this constant relies upon averaging over the region of $C_{ijkl}(\vec{q})$ where the magnitude of the function bears no dependence on $|q|$ (i.e., where the measured function agrees with the theoretical prediction for the $|q|$-dependence of the stress correlations). Although the angular dependence of the correlation functions measured for floppy gels are indeed consistent with the angular dependence prescribed by VCTG, the altered $|q|$-dependence precludes them from having a well-defined magnitude to fit to VCTG theoretical forms.

The results of this process are displayed in Fig. \ref{pressurestory}d) for all the rigid gels. The data show that the stiffness increases with pressure for rigid gels, and that there is a trend of the gel modulus increasing with pressure which becomes more clear at higher particle surface fraction. Overall the analysis performed confirms further the agreement with the VCTG theory, and how exploring the theory predictions can provide new insight into the mechanics of the gels, which indicate interesting directions for future investigations. For example: some signatures of the possible coupling between stiffness and pressure indicated here could be seen in the confinement effect on the modulus detected in experiments on agarose gels \cite{mao2016normal}; combining the different results in Fig.\ref{pressurestory} could allow to understand how structure or pressure heterogeneities may change the gel modulus, and more.

\subsection*{Debye-like Screening}
In this section, we go back to the length scale associated with the stress fluctuations and present in both rigid and floppy gels, shown in Fig. \ref{crosssection} (d-f). As discussed above, there is a characteristic length scale above which the VCTG predictions strictly apply to the gels, distinguishing clearly rigid from floppy configurations. Over distances smaller than this length scale, stress correlations behave similarly for rigid and floppy gels and the predictions of VCTG do not hold.
These findings raise the question of a possible screening length for the theory.  We recall that the VCTG framework corresponds to {\it dielectric} screening, due to bound charges as in electrostatics, but the charges are forces, and therefore, vectors. The polarization field $\hat{P}$ represents the pairwise interaction forces between gel particles or grains where the two equal and opposite forces make up the force dipole. Screening of the type emerging from mobile charges could 
produce a screening length. As in electrostatics, the action of the VCTG field theory can be extended to include a length scale:
\begin{equation}
    \mathcal{S} = \int d^d \boldsymbol{r} ~ g\sigma_{ij}\Lambda_{ijkl} \sigma_{kl} + \tilde{g}\Gamma_{jl}(\partial_i P_{ij} )(\partial_k P_{kl}) ~,
\end{equation}
where $\sigma_{ij} = E_{ij} + P_{ij}$, and $\nabla \times \nabla \times E =0$, $\Lambda_{ijkl}$ is the 4th rank tensor which gives rise to the emergent elastic modulus, and $\Gamma_{ij},g,\tilde{g}$ are additional coupling constants of the theory. The screening length scale, just from dimension counting, is $\xi = \sqrt{\frac{\tilde{g}}{g}}$. This scalar length scale would manifest itself in stress-stress correlations in a complicated way depending on the coupling constants $\Lambda$ and $\Gamma$. However, following Ref. \cite{Livne:2023aa}, we can identify $\xi$ from the pressure-pressure correlations $\langle P({\bf{q}})P(\bf{-q})\rangle$ in 2D, which are completely isotropic, leading to a change from $q^{0}$ dependence to a $q^{-2}$ dependence around $1/q \simeq \xi$. In Fig. \ref{lengthscale} we show that this indeed happens for all $\langle P({\bf{q}})P(\bf{-q})\rangle$ computed from the simulations for the three classes of ensembles: configurations with rigid clusters percolating in (i) 2 directions, (ii) 1 direction, and (iii) 0 directions. The single configuration at $\phi=0.5$ which did not percolate in either direction has been omitted because of lack of statistics. While only rigid gels can be seen to approach a flat $|q|$-dependence near $q=0$, all of these correlation functions display a change in power-law behavior as predicted by the theory with screening. Given that the theory describes well the rigidity of the gels, one could expect the screening length to be sensitive to it, however the crossover between the two behaviors seems to occur at approximately the same characteristic length scale, which is estimated around 10-12 particle diameters, for both rigid and non rigid gels. In SCT, the screening length $\xi$ can diverge with unbinding of defects in crystalline solids~\cite{Pretko_Radzihovsky}, and analogs of such transitions have been observed in the loss of amorphous materials rigidity\cite{Livne:2023aa}. For our finite size samples of model gels, the screening length $\xi$ obtained from VCTG does not seem to diverge with the loss of rigidity. Future studies with varying system sizes and gel models could help elucidate the origin of the screening and clarify its physical meaning in the context of VCTG theory.

\section*{Conclusion}
\label{conclu}
We have tested quantitatively the predictions of a vector charge theory (VCTG) for amorphous solids with computer simulations of 2D model colloidal gels. The stress correlations in the model colloidal gels feature precisely the pinch-point singularity predicted by the theory as a result of the constraints imposed by the mechanical equilibrium to the internal forces. We have demonstrated that the size of the pinch-point singularity allows to quantitatively distinguish rigid from floppy gels, and introduces a direct connection between the vector charge theory and the rigidity percolation framework previously introduced to describe colloidal gelation and rigidity transitions for a range of soft materials. The angular dependence of the stress correlations in Fourier space is also quantitatively predicted by the theory and allows for extracting the emergent elastic modulus of the gels. The striking agreement of the results obtained through the numerical simulations of the gel model with the theory predictions suggest that the theory description of pre-stress may adequately capture the different sources of pre-stresses in systems like colloidal gels. 

We used the analysis performed to investigate correlations between gel rigidity and distributions of internal pressures across gels obtained in similar conditions (same preparation protocol and particle density). The results suggest that, for the same densities, there is an interplay between the presence of a rigid spanning cluster and the heterogeneities of pressures in gel configurations, which could have interesting potential implications for the mechanical properties. The stress correlations computed in the model gels also reveal some deviations from the VCTG theory, indicating a length scale beyond which the VCTG theory applies and rigid/floppy states can be clearly distinguished. A Debye-like screening mechanism seems to be able to capture these deviations from the VCTG theory in terms of a screening length, suggesting that a screening of the type due to mobile charges could be limited to very short lengthscale (i.e. the grain size) in granular solids but extends over much larger lengthscales in self-assembled gels, pointing to a possible connection with geometric frustration in the gel structures. Future investigations of the screening length and of its microscopic origin across different gel models and self-assembly processes could provide more insight into these questions. 


\section*{Author contributions}
A.C. and H.A.V. performed all the analysis. F.D.R and H.A.V performed the numerical simulations of gels. A.C., H.A.V., X.M., E.D.G. and B.C. designed the research, discussed the interpretation and wrote the article. 

\section*{Conflicts of interest}
There are no conflicts to declare.


\section*{Acknowledgements}
We thank Jishnu N. Nampoothiri for valuable discussions. This work was supported by the National Science Foundation (NSF-DMR-2026842; NSF-DMR-2026825 and NSF-DMR-2026834).



\balance


\bibliography{forcechain_gels_v2,Response_Bibliography,granular_stress}

\providecommand*{\mcitethebibliography}{\thebibliography}
\csname @ifundefined\endcsname{endmcitethebibliography}
{\let\endmcitethebibliography\endthebibliography}{}
\begin{mcitethebibliography}{43}
\providecommand*{\natexlab}[1]{#1}
\providecommand*{\mciteSetBstSublistMode}[1]{}
\providecommand*{\mciteSetBstMaxWidthForm}[2]{}
\providecommand*{\mciteBstWouldAddEndPuncttrue}
  {\def\EndOfBibitem{\unskip.}}
\providecommand*{\mciteBstWouldAddEndPunctfalse}
  {\let\EndOfBibitem\relax}
\providecommand*{\mciteSetBstMidEndSepPunct}[3]{}
\providecommand*{\mciteSetBstSublistLabelBeginEnd}[3]{}
\providecommand*{\EndOfBibitem}{}
\mciteSetBstSublistMode{f}
\mciteSetBstMaxWidthForm{subitem}
{(\emph{\alph{mcitesubitemcount}})}
\mciteSetBstSublistLabelBeginEnd{\mcitemaxwidthsubitemform\space}
{\relax}{\relax}

\bibitem[Del~Gado \emph{et~al.}(2016)Del~Gado, Fiocco, Foffi, Manley, Trappe, and Zaccone]{DelGadoBookChapter2016}
E.~Del~Gado, D.~Fiocco, G.~Foffi, S.~Manley, V.~Trappe and A.~Zaccone, in \emph{Colloidal Gelation}, John Wiley \& Sons, Ltd, 2016, ch.~14, pp. 279--291\relax
\mciteBstWouldAddEndPuncttrue
\mciteSetBstMidEndSepPunct{\mcitedefaultmidpunct}
{\mcitedefaultendpunct}{\mcitedefaultseppunct}\relax
\EndOfBibitem
\bibitem[Tadros(2015)]{tadros2015interfacial}
T.~F. Tadros, \emph{Interfacial Phenomena and Colloid Stability: Industrial Applications}, Walter de Gruyter GmbH \& Co KG, 2015\relax
\mciteBstWouldAddEndPuncttrue
\mciteSetBstMidEndSepPunct{\mcitedefaultmidpunct}
{\mcitedefaultendpunct}{\mcitedefaultseppunct}\relax
\EndOfBibitem
\bibitem[Petekidis and Wagner(2021)]{petekidis_wagner_2021}
G.~Petekidis and N.~J. Wagner, in \emph{Rheology of Colloidal Glasses and Gels}, Cambridge University Press, 2021, p. 173–226\relax
\mciteBstWouldAddEndPuncttrue
\mciteSetBstMidEndSepPunct{\mcitedefaultmidpunct}
{\mcitedefaultendpunct}{\mcitedefaultseppunct}\relax
\EndOfBibitem
\bibitem[Cao and Mezzenga(2020)]{cao2020design}
Y.~Cao and R.~Mezzenga, \emph{Nature Food}, 2020, \textbf{1}, 106--118\relax
\mciteBstWouldAddEndPuncttrue
\mciteSetBstMidEndSepPunct{\mcitedefaultmidpunct}
{\mcitedefaultendpunct}{\mcitedefaultseppunct}\relax
\EndOfBibitem
\bibitem[Aime \emph{et~al.}(2018)Aime, Ramos, and Cipelletti]{aime2018microscopic}
S.~Aime, L.~Ramos and L.~Cipelletti, \emph{Proceedings of the National Academy of Sciences}, 2018, \textbf{115}, 3587--3592\relax
\mciteBstWouldAddEndPuncttrue
\mciteSetBstMidEndSepPunct{\mcitedefaultmidpunct}
{\mcitedefaultendpunct}{\mcitedefaultseppunct}\relax
\EndOfBibitem
\bibitem[Filiberti \emph{et~al.}(2019)Filiberti, Piazza, and Buzzaccaro]{filiberti2019multiscale}
Z.~Filiberti, R.~Piazza and S.~Buzzaccaro, \emph{Phys. Rev. E}, 2019, \textbf{100}, 042607\relax
\mciteBstWouldAddEndPuncttrue
\mciteSetBstMidEndSepPunct{\mcitedefaultmidpunct}
{\mcitedefaultendpunct}{\mcitedefaultseppunct}\relax
\EndOfBibitem
\bibitem[Colombo and Del~Gado(2014)]{colombo2014stress}
J.~Colombo and E.~Del~Gado, \emph{Journal of rheology}, 2014, \textbf{58}, 1089--1116\relax
\mciteBstWouldAddEndPuncttrue
\mciteSetBstMidEndSepPunct{\mcitedefaultmidpunct}
{\mcitedefaultendpunct}{\mcitedefaultseppunct}\relax
\EndOfBibitem
\bibitem[Koumakis \emph{et~al.}(2015)Koumakis, Moghimi, Besseling, Poon, Brady, and Petekidis]{koumakis2015tuning}
N.~Koumakis, E.~Moghimi, R.~Besseling, W.~C. Poon, J.~F. Brady and G.~Petekidis, \emph{Soft Matter}, 2015, \textbf{11}, 4640--4648\relax
\mciteBstWouldAddEndPuncttrue
\mciteSetBstMidEndSepPunct{\mcitedefaultmidpunct}
{\mcitedefaultendpunct}{\mcitedefaultseppunct}\relax
\EndOfBibitem
\bibitem[Gibaud \emph{et~al.}(2022)Gibaud, Divoux, and Manneville]{gibaud2022nonlinear}
T.~Gibaud, T.~Divoux and S.~Manneville, \emph{Statistical and Nonlinear Physics}, Springer, 2022, pp. 313--336\relax
\mciteBstWouldAddEndPuncttrue
\mciteSetBstMidEndSepPunct{\mcitedefaultmidpunct}
{\mcitedefaultendpunct}{\mcitedefaultseppunct}\relax
\EndOfBibitem
\bibitem[Alexander(1998)]{alexander1998amorphous}
S.~Alexander, \emph{Physics reports}, 1998, \textbf{296}, 65--236\relax
\mciteBstWouldAddEndPuncttrue
\mciteSetBstMidEndSepPunct{\mcitedefaultmidpunct}
{\mcitedefaultendpunct}{\mcitedefaultseppunct}\relax
\EndOfBibitem
\bibitem[Anderson and Lekkerkerker(2002)]{anderson2002insights}
V.~J. Anderson and H.~N. Lekkerkerker, \emph{Nature}, 2002, \textbf{416}, 811--815\relax
\mciteBstWouldAddEndPuncttrue
\mciteSetBstMidEndSepPunct{\mcitedefaultmidpunct}
{\mcitedefaultendpunct}{\mcitedefaultseppunct}\relax
\EndOfBibitem
\bibitem[Royall \emph{et~al.}(2021)Royall, Faers, Fussell, and Hallett]{royall2021real}
C.~P. Royall, M.~A. Faers, S.~L. Fussell and J.~E. Hallett, \emph{J. Condens. Matter Phys.}, 2021, \textbf{33}, 453002\relax
\mciteBstWouldAddEndPuncttrue
\mciteSetBstMidEndSepPunct{\mcitedefaultmidpunct}
{\mcitedefaultendpunct}{\mcitedefaultseppunct}\relax
\EndOfBibitem
\bibitem[Mao \emph{et~al.}(2009)Mao, Goldbart, Xing, and Zippelius]{mao2009soft}
X.~Mao, P.~M. Goldbart, X.~Xing and A.~Zippelius, \emph{Phys. Rev. E}, 2009, \textbf{80}, 031140\relax
\mciteBstWouldAddEndPuncttrue
\mciteSetBstMidEndSepPunct{\mcitedefaultmidpunct}
{\mcitedefaultendpunct}{\mcitedefaultseppunct}\relax
\EndOfBibitem
\bibitem[Divoux \emph{et~al.}(2013)Divoux, Grenard, and Manneville]{divoux2013rheological}
T.~Divoux, V.~Grenard and S.~Manneville, \emph{Physical review letters}, 2013, \textbf{110}, 018304\relax
\mciteBstWouldAddEndPuncttrue
\mciteSetBstMidEndSepPunct{\mcitedefaultmidpunct}
{\mcitedefaultendpunct}{\mcitedefaultseppunct}\relax
\EndOfBibitem
\bibitem[Bouzid \emph{et~al.}(2017)Bouzid, Colombo, Barbosa, and Del~Gado]{bouzid2017elastically}
M.~Bouzid, J.~Colombo, L.~V. Barbosa and E.~Del~Gado, \emph{Nature communications}, 2017, \textbf{8}, 1--8\relax
\mciteBstWouldAddEndPuncttrue
\mciteSetBstMidEndSepPunct{\mcitedefaultmidpunct}
{\mcitedefaultendpunct}{\mcitedefaultseppunct}\relax
\EndOfBibitem
\bibitem[Behringer and Chakraborty(2018)]{behringer2018physics}
R.~P. Behringer and B.~Chakraborty, \emph{Reports on Progress in Physics}, 2018, \textbf{82}, 012601\relax
\mciteBstWouldAddEndPuncttrue
\mciteSetBstMidEndSepPunct{\mcitedefaultmidpunct}
{\mcitedefaultendpunct}{\mcitedefaultseppunct}\relax
\EndOfBibitem
\bibitem[Henkes and Schwarz(2022)]{henkes2022rigidity}
S.~Henkes and J.~Schwarz, \emph{Statistical and Nonlinear Physics}, Springer, 2022, pp. 427--448\relax
\mciteBstWouldAddEndPuncttrue
\mciteSetBstMidEndSepPunct{\mcitedefaultmidpunct}
{\mcitedefaultendpunct}{\mcitedefaultseppunct}\relax
\EndOfBibitem
\bibitem[Vinutha and Sastry(2019)]{vinutha2019force}
H.~A. Vinutha and S.~Sastry, \emph{Physical Review E}, 2019, \textbf{99}, 012123\relax
\mciteBstWouldAddEndPuncttrue
\mciteSetBstMidEndSepPunct{\mcitedefaultmidpunct}
{\mcitedefaultendpunct}{\mcitedefaultseppunct}\relax
\EndOfBibitem
\bibitem[Zhang \emph{et~al.}(2019)Zhang, Zhang, Bouzid, Rocklin, Del~Gado, and Mao]{zhang2019correlated}
S.~Zhang, L.~Zhang, M.~Bouzid, D.~Z. Rocklin, E.~Del~Gado and X.~Mao, \emph{Physical review letters}, 2019, \textbf{123}, 058001\relax
\mciteBstWouldAddEndPuncttrue
\mciteSetBstMidEndSepPunct{\mcitedefaultmidpunct}
{\mcitedefaultendpunct}{\mcitedefaultseppunct}\relax
\EndOfBibitem
\bibitem[Tong \emph{et~al.}(2020)Tong, Sengupta, and Tanaka]{tong2020emergent}
H.~Tong, S.~Sengupta and H.~Tanaka, \emph{Nature communications}, 2020, \textbf{11}, 4863\relax
\mciteBstWouldAddEndPuncttrue
\mciteSetBstMidEndSepPunct{\mcitedefaultmidpunct}
{\mcitedefaultendpunct}{\mcitedefaultseppunct}\relax
\EndOfBibitem
\bibitem[Dashti \emph{et~al.}(2023)Dashti, Saberi, Rahbari, and Kurths]{dashti2023emergence}
H.~Dashti, A.~A. Saberi, S.~Rahbari and J.~u.~r. Kurths, \emph{Science Advances}, 2023, \textbf{9}, eadh5586\relax
\mciteBstWouldAddEndPuncttrue
\mciteSetBstMidEndSepPunct{\mcitedefaultmidpunct}
{\mcitedefaultendpunct}{\mcitedefaultseppunct}\relax
\EndOfBibitem
\bibitem[Nampoothiri \emph{et~al.}(2020)Nampoothiri, Wang, Ramola, Zhang, Bhattacharjee, and Chakraborty]{nampoothiri2020emergent}
J.~N. Nampoothiri, Y.~Wang, K.~Ramola, J.~Zhang, S.~Bhattacharjee and B.~Chakraborty, \emph{Physical review letters}, 2020, \textbf{125}, 118002\relax
\mciteBstWouldAddEndPuncttrue
\mciteSetBstMidEndSepPunct{\mcitedefaultmidpunct}
{\mcitedefaultendpunct}{\mcitedefaultseppunct}\relax
\EndOfBibitem
\bibitem[Nampoothiri \emph{et~al.}(2022)Nampoothiri, D'Eon, Ramola, Chakraborty, and Bhattacharjee]{nampoothiri2022tensor}
J.~N. Nampoothiri, M.~D'Eon, K.~Ramola, B.~Chakraborty and S.~Bhattacharjee, \emph{arXiv preprint arXiv:2204.11811}, 2022\relax
\mciteBstWouldAddEndPuncttrue
\mciteSetBstMidEndSepPunct{\mcitedefaultmidpunct}
{\mcitedefaultendpunct}{\mcitedefaultseppunct}\relax
\EndOfBibitem
\bibitem[Pretko(2017)]{pretko2017generalized}
M.~Pretko, \emph{Physical Review B}, 2017, \textbf{96}, 035119\relax
\mciteBstWouldAddEndPuncttrue
\mciteSetBstMidEndSepPunct{\mcitedefaultmidpunct}
{\mcitedefaultendpunct}{\mcitedefaultseppunct}\relax
\EndOfBibitem
\bibitem[Whitaker \emph{et~al.}(2019)Whitaker, Varga, Hsiao, Solomon, Swan, and Furst]{whitaker2019colloidal}
K.~A. Whitaker, Z.~Varga, L.~C. Hsiao, M.~J. Solomon, J.~W. Swan and E.~M. Furst, \emph{Nature communications}, 2019, \textbf{10}, 2237\relax
\mciteBstWouldAddEndPuncttrue
\mciteSetBstMidEndSepPunct{\mcitedefaultmidpunct}
{\mcitedefaultendpunct}{\mcitedefaultseppunct}\relax
\EndOfBibitem
\bibitem[Keshavarz \emph{et~al.}(2021)Keshavarz, Rodrigues, Champenois, Frith, Ilavsky, Geri, Divoux, McKinley, and Poulesquen]{keshavarz2021time}
B.~Keshavarz, D.~G. Rodrigues, J.-B. Champenois, M.~G. Frith, J.~Ilavsky, M.~Geri, T.~Divoux, G.~H. McKinley and A.~Poulesquen, \emph{Proceedings of the National Academy of Sciences}, 2021, \textbf{118}, e2022339118\relax
\mciteBstWouldAddEndPuncttrue
\mciteSetBstMidEndSepPunct{\mcitedefaultmidpunct}
{\mcitedefaultendpunct}{\mcitedefaultseppunct}\relax
\EndOfBibitem
\bibitem[Hsiao \emph{et~al.}(2012)Hsiao, Newman, Glotzer, and Solomon]{hsiao2012role}
L.~C. Hsiao, R.~S. Newman, S.~C. Glotzer and M.~J. Solomon, \emph{Proceedings of the National Academy of Sciences}, 2012, \textbf{109}, 16029--16034\relax
\mciteBstWouldAddEndPuncttrue
\mciteSetBstMidEndSepPunct{\mcitedefaultmidpunct}
{\mcitedefaultendpunct}{\mcitedefaultseppunct}\relax
\EndOfBibitem
\bibitem[Bantawa \emph{et~al.}(2023)Bantawa, Keshavarz, Geri, Bouzid, Divoux, McKinley, and Del~Gado]{bantawa2023hidden}
M.~Bantawa, B.~Keshavarz, M.~Geri, M.~Bouzid, T.~Divoux, G.~H. McKinley and E.~Del~Gado, \emph{Nature Physics}, 2023, \textbf{19}, 1178--1184\relax
\mciteBstWouldAddEndPuncttrue
\mciteSetBstMidEndSepPunct{\mcitedefaultmidpunct}
{\mcitedefaultendpunct}{\mcitedefaultseppunct}\relax
\EndOfBibitem
\bibitem[Fenton \emph{et~al.}(2023)Fenton, Padmanabhan, Ryu, Nguyen, Zia, and Helgeson]{fenton2023minimal}
S.~M. Fenton, P.~Padmanabhan, B.~K. Ryu, T.~T. Nguyen, R.~N. Zia and M.~E. Helgeson, \emph{Proceedings of the National Academy of Sciences}, 2023, \textbf{120}, e2215922120\relax
\mciteBstWouldAddEndPuncttrue
\mciteSetBstMidEndSepPunct{\mcitedefaultmidpunct}
{\mcitedefaultendpunct}{\mcitedefaultseppunct}\relax
\EndOfBibitem
\bibitem[Vinutha \emph{et~al.}(2023)Vinutha, Diaz~Ruiz, Mao, Chakraborty, and Del~Gado]{vinutha2023stress}
H.~A. Vinutha, F.~D. Diaz~Ruiz, X.~Mao, B.~Chakraborty and E.~Del~Gado, \emph{The Journal of chemical physics}, 2023, \textbf{158}, \relax
\mciteBstWouldAddEndPuncttrue
\mciteSetBstMidEndSepPunct{\mcitedefaultmidpunct}
{\mcitedefaultendpunct}{\mcitedefaultseppunct}\relax
\EndOfBibitem
\bibitem[Livne \emph{et~al.}(2023)Livne, Schiller, and Moshe]{Livne:2023aa}
N.~S. Livne, A.~Schiller and M.~Moshe, \emph{Phys Rev E}, 2023, \textbf{107}, 055004\relax
\mciteBstWouldAddEndPuncttrue
\mciteSetBstMidEndSepPunct{\mcitedefaultmidpunct}
{\mcitedefaultendpunct}{\mcitedefaultseppunct}\relax
\EndOfBibitem
\bibitem[Lema{\^\i}tre \emph{et~al.}(2021)Lema{\^\i}tre, Mondal, Moshe, Procaccia, Roy, and Screiber-Re'em]{lemaitre2021anomalous}
A.~Lema{\^\i}tre, C.~Mondal, M.~Moshe, I.~Procaccia, S.~Roy and K.~Screiber-Re'em, \emph{Physical Review E}, 2021, \textbf{104}, 024904\relax
\mciteBstWouldAddEndPuncttrue
\mciteSetBstMidEndSepPunct{\mcitedefaultmidpunct}
{\mcitedefaultendpunct}{\mcitedefaultseppunct}\relax
\EndOfBibitem
\bibitem[Kumar \emph{et~al.}(2022)Kumar, Moshe, Procaccia, and Singh]{Kumar:2022aa}
A.~Kumar, M.~Moshe, I.~Procaccia and M.~Singh, \emph{Phys Rev E}, 2022, \textbf{106}, 015001\relax
\mciteBstWouldAddEndPuncttrue
\mciteSetBstMidEndSepPunct{\mcitedefaultmidpunct}
{\mcitedefaultendpunct}{\mcitedefaultseppunct}\relax
\EndOfBibitem
\bibitem[Fu \emph{et~al.}(2024)Fu, Jin, Pan, and Procaccia]{fu2024long}
Y.~Fu, Y.~Jin, D.~Pan and I.~Procaccia, \emph{arXiv preprint arXiv:2410.04138}, 2024\relax
\mciteBstWouldAddEndPuncttrue
\mciteSetBstMidEndSepPunct{\mcitedefaultmidpunct}
{\mcitedefaultendpunct}{\mcitedefaultseppunct}\relax
\EndOfBibitem
\bibitem[Chakraborty \emph{et~al.}()Chakraborty, Nampoothiri, Bhattacharjee, Chakraborty, and Ramola]{Surajit_Kabir}
S.~Chakraborty, J.~N. Nampoothiri, S.~Bhattacharjee, B.~Chakraborty and K.~Ramola, In Preparation\relax
\mciteBstWouldAddEndPuncttrue
\mciteSetBstMidEndSepPunct{\mcitedefaultmidpunct}
{\mcitedefaultendpunct}{\mcitedefaultseppunct}\relax
\EndOfBibitem
\bibitem[Pretko \emph{et~al.}(2019)Pretko, Zhai, and Radzihovsky]{Pretko_Radzihovsky}
M.~Pretko, Z.~Zhai and L.~Radzihovsky, \emph{Phys. Rev. B}, 2019, \textbf{100}, 134113\relax
\mciteBstWouldAddEndPuncttrue
\mciteSetBstMidEndSepPunct{\mcitedefaultmidpunct}
{\mcitedefaultendpunct}{\mcitedefaultseppunct}\relax
\EndOfBibitem
\bibitem[Balents(2016)]{spin_liquids_review}
L.~Balents, \emph{Nature}, 2016, \textbf{540}, 534--535\relax
\mciteBstWouldAddEndPuncttrue
\mciteSetBstMidEndSepPunct{\mcitedefaultmidpunct}
{\mcitedefaultendpunct}{\mcitedefaultseppunct}\relax
\EndOfBibitem
\bibitem[Noro and Frenkel(2000)]{noro2000extended}
M.~G. Noro and D.~Frenkel, \emph{The Journal of Chemical Physics}, 2000, \textbf{113}, 2941--2944\relax
\mciteBstWouldAddEndPuncttrue
\mciteSetBstMidEndSepPunct{\mcitedefaultmidpunct}
{\mcitedefaultendpunct}{\mcitedefaultseppunct}\relax
\EndOfBibitem
\bibitem[Colombo and Del~Gado(2014)]{colombo2014self}
J.~Colombo and E.~Del~Gado, \emph{Soft Matter}, 2014, \textbf{10}, 4003--4015\relax
\mciteBstWouldAddEndPuncttrue
\mciteSetBstMidEndSepPunct{\mcitedefaultmidpunct}
{\mcitedefaultendpunct}{\mcitedefaultseppunct}\relax
\EndOfBibitem
\bibitem[Bantawa \emph{et~al.}(2021)Bantawa, Fontaine-Seiler, Olmsted, and Del~Gado]{bantawa2021microscopic}
M.~Bantawa, W.~A. Fontaine-Seiler, P.~D. Olmsted and E.~Del~Gado, \emph{Journal of Physics: Condensed Matter}, 2021, \textbf{33}, 414001\relax
\mciteBstWouldAddEndPuncttrue
\mciteSetBstMidEndSepPunct{\mcitedefaultmidpunct}
{\mcitedefaultendpunct}{\mcitedefaultseppunct}\relax
\EndOfBibitem
\bibitem[Plimpton(1995)]{lammps}
S.~Plimpton, \emph{Journal of computational physics}, 1995, \textbf{117}, 1--19\relax
\mciteBstWouldAddEndPuncttrue
\mciteSetBstMidEndSepPunct{\mcitedefaultmidpunct}
{\mcitedefaultendpunct}{\mcitedefaultseppunct}\relax
\EndOfBibitem
\bibitem[Jacobs and Hendrickson(1997)]{jacobs1997algorithm}
D.~J. Jacobs and B.~Hendrickson, \emph{Journal of Computational Physics}, 1997, \textbf{137}, 346--365\relax
\mciteBstWouldAddEndPuncttrue
\mciteSetBstMidEndSepPunct{\mcitedefaultmidpunct}
{\mcitedefaultendpunct}{\mcitedefaultseppunct}\relax
\EndOfBibitem
\bibitem[Mao \emph{et~al.}(2016)Mao, Divoux, and Snabre]{mao2016normal}
B.~Mao, T.~Divoux and P.~Snabre, \emph{Journal of Rheology}, 2016, \textbf{60}, 473--489\relax
\mciteBstWouldAddEndPuncttrue
\mciteSetBstMidEndSepPunct{\mcitedefaultmidpunct}
{\mcitedefaultendpunct}{\mcitedefaultseppunct}\relax
\EndOfBibitem
\end{mcitethebibliography}
\bibliographystyle{rsc} 

\end{document}